\documentclass[prb,twocolumn,showpacs,preprintnumbers,amsmath,amssymb,floatfix]{revtex4}

\usepackage{graphicx}
\usepackage{dcolumn}
\usepackage{bm}

\newcommand{\Tlowlevel}[1]{(TMTTF)$_2$#1}
\newcommand{\Clowlevel}[1]{(TMT$C$F)$_2$#1}
\newcommand{\Slowlevel}[1]{(TMTSF)$_2$#1}
\newcommand{\TAsF}{\Tlowlevel{AsF$_6$}}           
\newcommand{\CX}{\Clowlevel{$X$}}
\newcommand{\SX}{\Slowlevel{$X$}}
\newcommand{\TX}{\Tlowlevel{$X$}}
\newcommand{\PF}{\Slowlevel{PF$_6$}}                 
\newcommand{\myreffig}[1]{Fig.\ \ref{#1}}

\newcommand{\Omcminv}{$\,(\Omega\text{cm})^{-1}$}

\newcommand{\mTCO}{T_{\mathrm{CO}}}     \newcommand{\TCO}{$\mTCO$}
\newcommand{\parl}[2]{\mv{#1}\|\mv{#2}}

\newcommand{\mv}[1]{\mathbf{#1}}               
\newcommand{\vek}[1]{$\mv{#1}$}                

\newcommand{\mva}{\mv{a}}   \newcommand{\va}{$\mva$}
\newcommand{\mvb}{\mv{b'}}  \newcommand{\vb}{$\mvb$}
\newcommand{\mvc}{\mv{c^\star}}   \newcommand{\vc}{$\mvc$}

\newcommand{\mja}{\parl{j}{a}}          \newcommand{\ja}{$\mja$}
\newcommand{\mjb}{\parl{j}{b'}}         \newcommand{\jb}{$\mjb$}
\newcommand{\mjc}{\parl{j}{c^\star}}    \newcommand{\jc}{$\mjc$}

\newcommand{\mBc}{\parl{B}{c^\star}}    \newcommand{\Bc}{$\mBc$}

\newcommand{\rhoa}{\rho_\mva}
\newcommand{\rhob}{\rho_\mvb}
\newcommand{\rhoc}{\rho_\mvc}

\newcommand{\plane}[2]{#1\text{--}#2}
\newcommand{\macplane}{\plane{\mva}{\mvc}}
\newcommand{\acplane}{$\macplane$}
\newcommand{\mabplane}{\plane{\mva}{\mvb}}
\newcommand{\abplane}{$\mabplane$}

\begin{document}

\title{Conduction anisotropy and Hall effect in the organic conductor (TMTTF)$_2$AsF$_6$:
    evidence for Luttinger liquid and charge ordering}

\author{Bojana Korin-Hamzi\'c}
    \email{bhamzic@ifs.hr}
    \affiliation{Institute of Physics, P.O.Box 304, HR-10001 Zagreb, Croatia}

\author{Emil Tafra, Mario Basleti\'c and Amir Hamzi\'c}
    \affiliation{Department of Physics, Faculty of Science, P.O.Box 331, HR-10002 Zagreb, Croatia}

\author{Martin Dressel}
    \affiliation{1.~Physikalisches Institut, Universit\"{a}t Stuttgart, Pfaffenwaldring 57, D-70550 Stuttgart, Germany}
\date{\today}

\begin{abstract}
We present the high-temperature
($70\,\mathrm{K}<T<300\,\mathrm{K}$) resistivity anisotropy and
Hall effect measurements of the quasi-one-dimensional (1D) organic
conductor (TMTTF)$_2$AsF$_6$. The temperature variations of the
resistivity are pronouncedly different for the three different
directions, with metallic-like at high temperatures for the
$\mathbf{a}$-axis only. Above $220\,$K the Hall coefficient $R_H$
is constant, positive and strongly enhanced over the expected
value; and the corresponding carrier concentration is almost 100
times lower than calculated for one hole/unit cell. Our results
give evidence for the existence of a high-temperature regime above
$200\,$K where the 1D Luttinger liquid features appear in the
transport properties. Our measurements also give strong evidence
of charge ordering in (TMTTF)$_2$AsF$_6$. At the charge-ordering
transition $T_{\mathrm{CO}}\approx 100\,$K, $R_H(T)$ abruptly
changes its behavior, switches sign and rapidly increases with
further temperature decrease.

\end{abstract}


\maketitle

\section{Introduction}
The highly anisotropic organic conductors with their very rich phase diagram
turn out to be excellent choice for exploring the phenomena of one-dimensional
physics.\cite{IshiguroBook98-1} It is well established theoretically that the
conventional Fermi-liquid (FL) theory of 3D metals does not apply to
interacting electrons when the motion is confined to one dimension. Instead,
the Luttinger liquid (LL) approach is valid, where the quasi-particle FL
excitations are replaced by separate collective spin and charge excitations,
propagating with different velocities.\cite{Schulz91-2,Voit92-3} LL systems
exhibit non-FL like power-law behavior in temperature and energy, and the
exponents are interaction-dependent. When coupling between the chains is
relevant, a crossover from the LL to a coherent (FL) behavior is expected as
the temperature (or frequency) is lowered.\cite{Giamarchi04}

The quasi-1D character of the electronic structure of \CX\ ($C=\mathrm{Se}$,
tetramethyltetraselenafulvalene; $C=\mathrm{S}$, tetramethyltetrathiofulvalene;
anion $X^-=\,\,$PF$_6^-$, AsF$_6^-$, ReO$_4^-$, \ldots) materials is a
consequence of their crystallographic structure. The large planar TMT$C$F
molecules form segregated molecular columns (chains) where the molecular
orbitals overlap along the columns, giving rise to the highest conductivity
direction (\vek{a}). The separation of the columns limits the coupling in the
perpendicular directions. The anisotropy of the tight-binding overlap integrals
along the three directions is $t_a:t_b:t_c\approx 1:0.1:0.005$ for the high
(\vek{a}), intermediate (\vek{b}) and weak (\vek{c}) coupling/conductivity
directions respectively.\cite{IshiguroBook98-1} Charge transfer of one electron
from the chains to each of the anions results in quarter-filled hole bands on
the TMT$C$F stacks (or half-filled, if the dimerization along the chain is
taken into account). It is also well known that exchanging Se for S and/or
using a different anion $X$ leads to the unified phase diagram for the \CX\
series that spans from more anisotropic $C=\mathrm{S}$ to $C=\mathrm{Se}$,
where the properties of one compound at a given pressure are analogous to those
of another compound under higher pressure.\cite{Jerome94-4}

Selenium based \SX\ salts exhibit a high conductivity at room temperature and a
metallic behavior down to low temperatures, where the incommensurate
spin-density-wave (SDW) ground state is established. Like other $C=\mathrm{S}$
salts that are more anisotropic, i.e.\ more 1D, \TAsF\ is typically 20 (or
more) times less conducting than $C=\mathrm{Se}$ salts at room
tem\-pe\-ra\-tu\-re.\cite{Laversanne84-5,Bourbonnais99-6} Below a broad minimum
near $T_\rho \approx 240\,$K (attributed to the opening of a charge gap
$\Delta_\rho$ and considered to be a continuous charge localization due to the
anion potential -- lattice dimerization), the resistivity of \TAsF\ increases
on cooling and the system becomes a very good insulator at low temperatures
($T<20\,$K), where the spin-Peierls transition occurs. The spin susceptibility
remains unaffected by the charge localization despite the thermal activation of
carriers below $T_\rho$, indicating the decoupling between the spin and charge
degrees of freedom. The anomaly in resistivity at $\mTCO\approx 100\,$K has
been recently clarified by the discoveries of the huge anomaly in the
dielectric constant $\varepsilon$ (Ref.\ \onlinecite{Monceau01-7}) and of the
charge ordering (CO) seen by nuclear magnetic resonance (NMR), electron spin
resonance (ESR) and optical
experiments.\cite{Chow00-8,Dumm04,Nakamura03,Salameh05} The nature of this
phase transition was interpreted as a ferroelectric (FE) state, which is
supported by the clear-cut fit of the anomaly in $\varepsilon (T)$ to a Curie
law. The FE transition is followed by a steep increase of the conductivity gap
$\Delta$ but with no appearance of a spin gap.

There is still no general agreement whether the transport properties of
quasi-1D \CX\ salts at high temperatures ($T>200\,$K) -- where the thermal
energy exceeds the transverse coupling and the coherence for the interchain
transport is lost -- should be understood in terms of usual FL theory or LL
theory.\cite{Bourbonnais99-6,Dressel03-9} For example, there are indications of
spin-charge separation by the similarity in the spin dynamics and thermal
conductivity for $C=\mathrm{S}$ and $C=\mathrm{Se}$ salts, but the electronic
transport is very different.

Further, the absence of a plasma edge for polarization perpendicular to the
chains in \TX\ salts suggests that the electrons are confined on the chains,
which is not the case for \SX\ compounds. However, while the dc conductivities
along the chain axis \vek{a} in the latter compounds are metallic-like, the
finite frequency response is not that of a simple Drude metal and the optical
conductivity data for \CX\ salts were interpreted as a strong evidence for
non-FL behavior: the power law asymptotic dependence of the high frequency
optical mode has been addressed to LL exponents.\cite{Vescoli98-10} It was
concluded that all the dc transport is due to a very narrow Drude peak,
containing only 1\% of the spectral weight, whereas the remaining 99\% is above
an energy gap (of the order of $200\,$cm$^{-1}$) and reminiscent of a Mott
insulating structure. These results agree with the LL theoretical predictions
based on the doped Hubbard chains, yielding a gap feature and a zero-frequency
mode with a small spectral weight (this mode is responsible for the metallic
conductivity).\cite{Giamarchi97-11} One would expect that such a reduction of
the carrier concentration $n$ (participating in the dc transport) should
manifest itself in an enhanced value of the Hall coefficient. However, so far
the high temperature Hall effect measurements were performed for \SX\ compounds
only, and did not confirm these expectations; moreover, the results for \PF\
were interpreted differently, i.e.\ using the conventional FL theory
\cite{Mihaly00-12} and LL concept.\cite{Moser00-13} Recently, we have
investigated the metallic state of (TMTSF)$_2$ReO$_4$ (more anisotropic than
the $X=\,\,$PF$_6$), aiming to shed more light on FL or LL concepts from
transport measurements and particularly Hall effect.\cite{Bojana03-14} The
pronounced conductivity anisotropy, a small and smoothly temperature dependent
Hall effect (where the obtained $R_H$ values  indicate that all the carriers
contribute to the metallic-like transport) and a small, positive and
temperature dependent magnetoresistance were analyzed within FL and non-FL
models: although not fully conclusive, these data nevertheless favor the FL
description. We have also proposed that the possible appearance of the LL
features in the transport properties should be found in the more anisotropic
\TX\ series, where the interchain interactions play a minor role. The present
study of the temperature dependence of the resistivity anisotropy and the Hall
effect in \TAsF\ is therefore a step forward in the identification of the LL
features in the transport properties of $C=\mathrm{S}$ compounds. In addition,
our study gives strong evidence for a charge ordering transition from transport
properties.

\section{Experiment}
The measurements were conducted in the high temperature region
$70\,\mathrm{K}<T<300\,\mathrm{K}$. The single crystals came all from the same
batch and exhibit the same behavior. The resistivity data, that will be
analyzed here, are for \va, \vb\ and \vc\ axis. The \va\ direction is the
highest conductivity direction, the \vb\ direction (with intermediate
conductivity) is perpendicular to \va\ in the $\plane{\mva}{\mv{b}}$ plane and
the \vc\ direction (with the lowest conductivity) is perpendicular to the
$\plane{\mva}{\mv{b}}$ (and \abplane) plane. For \vb\ and \vc-axis resistivity
measurements, the samples have been cut from a long crystal along \va\ axis,
and the contacts were placed on the opposite \acplane\ ($\rhob$) and \abplane\
($\rhoc$) surfaces with gold wires stuck by silver paint. The samples were
cooled slowly ($3\,$K/hour) in order to avoid irreversible resistance jumps
(caused by micro-cracks), well known to appear in all organic conductors.

The Hall effect was measured in standard geometry (\ja, \Bc) for temperatures
$70\,\mathrm{K}<T<300\,\mathrm{K}$ and in magnetic field up to $9\,$T. Two
pairs of Hall contacts and one pair of current contacts were made on the sides
of the crystal by evaporating gold pads and attaching $30\,\mu$m diameter gold
wires with silver paint.\cite{Cooper94-15} AC current ($10\,\mu$A to $1\,$mA,
$22\,$Hz) was applied. DC technique was used at lower temperatures, because of
the large resistance increment. Particular care was taken to ensure the
temperature stabilization. The Hall voltage was measured at fixed temperatures
for both pairs of Hall contacts to test and/or control the homogenous current
distribution through the sample and in field sweeps from $-B_{\mathrm{max}}$ to
$+B_{\mathrm{max}}$  in order to eliminate the possible mixing of
magnetoresistive components. The Hall coefficient $R_H$ was obtained as $R_H =
(V_{xy}/IB)t$, where $V_{xy}$ is Hall voltage determined as
$[V_{xy}(B)-V_{xy}(-B)]/2$, $I$ is the current through the crystal and $t$ is
the sample thickness.

\section{Results}
\begin{figure}
\includegraphics*[scale=0.58]{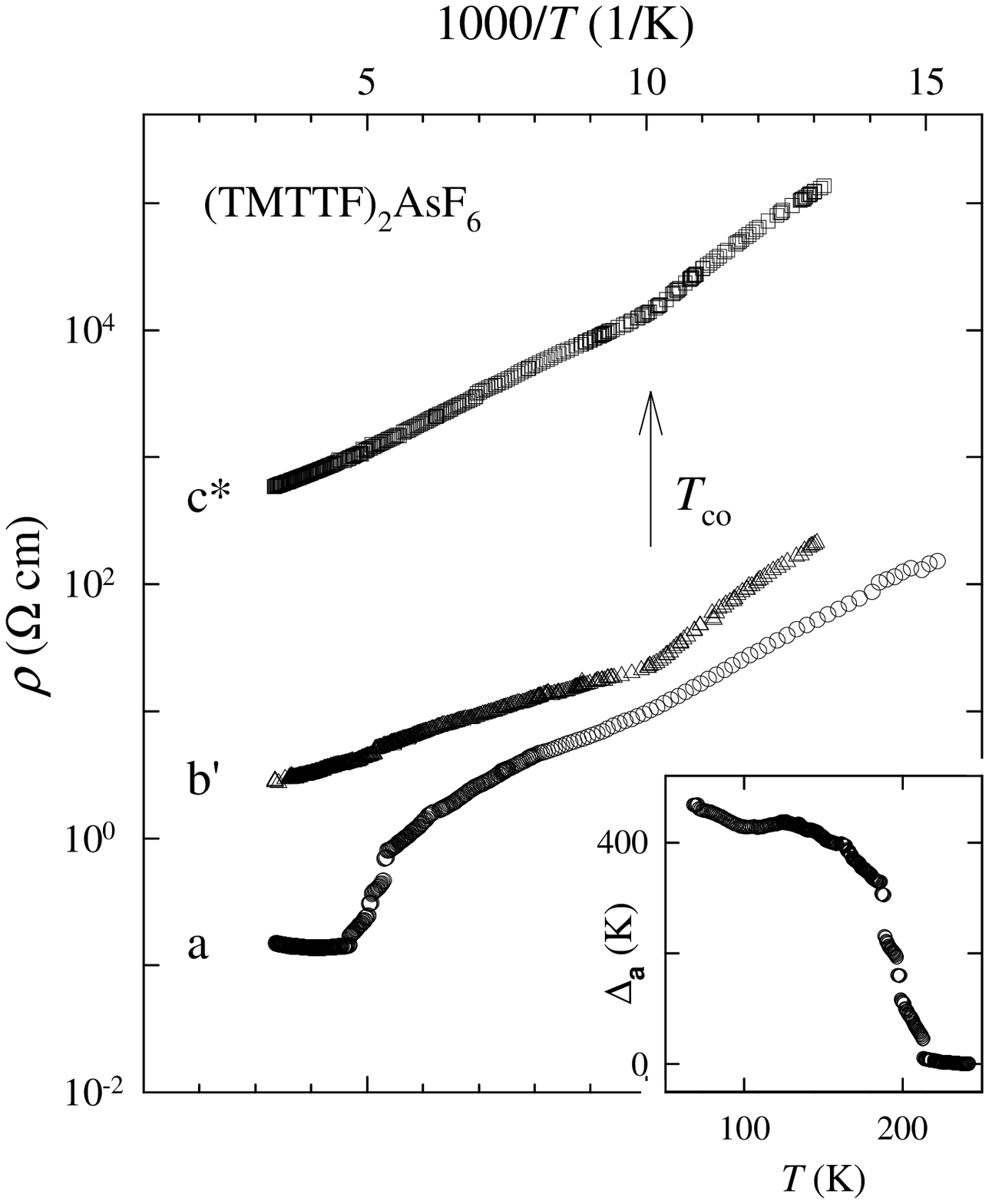}
\caption{\label{fig:2} Temperature dependence of the dc resistivity, $\rho$
vs.\ $1/T$, for \TAsF\ in the temperature range
$70\,\mathrm{K}<T<300\,\mathrm{K}$, measured along the three crystal
directions: $\rhoa$ (\ja), $\rhob$ (\jb) and $\rhoc$ (\jc). Inset: Activation
energy $\Delta_\mva$ vs.\ temperature calculated for the \va-axis assuming
$\Delta_\mva = k_BT\ln(\rhoa/\rho_{\mathrm{min}})$ - see text.}
\end{figure}

Figure \ref{fig:2} shows the temperature dependence of the resistivity,
$\rho(T)$ vs.\ $1/T$, for \TAsF\ in the temperature range
$70\,\mathrm{K}<T<300\,\mathrm{K}$, measured along the three crystal
directions. The
\begin{figure}
\includegraphics*[scale=0.58]{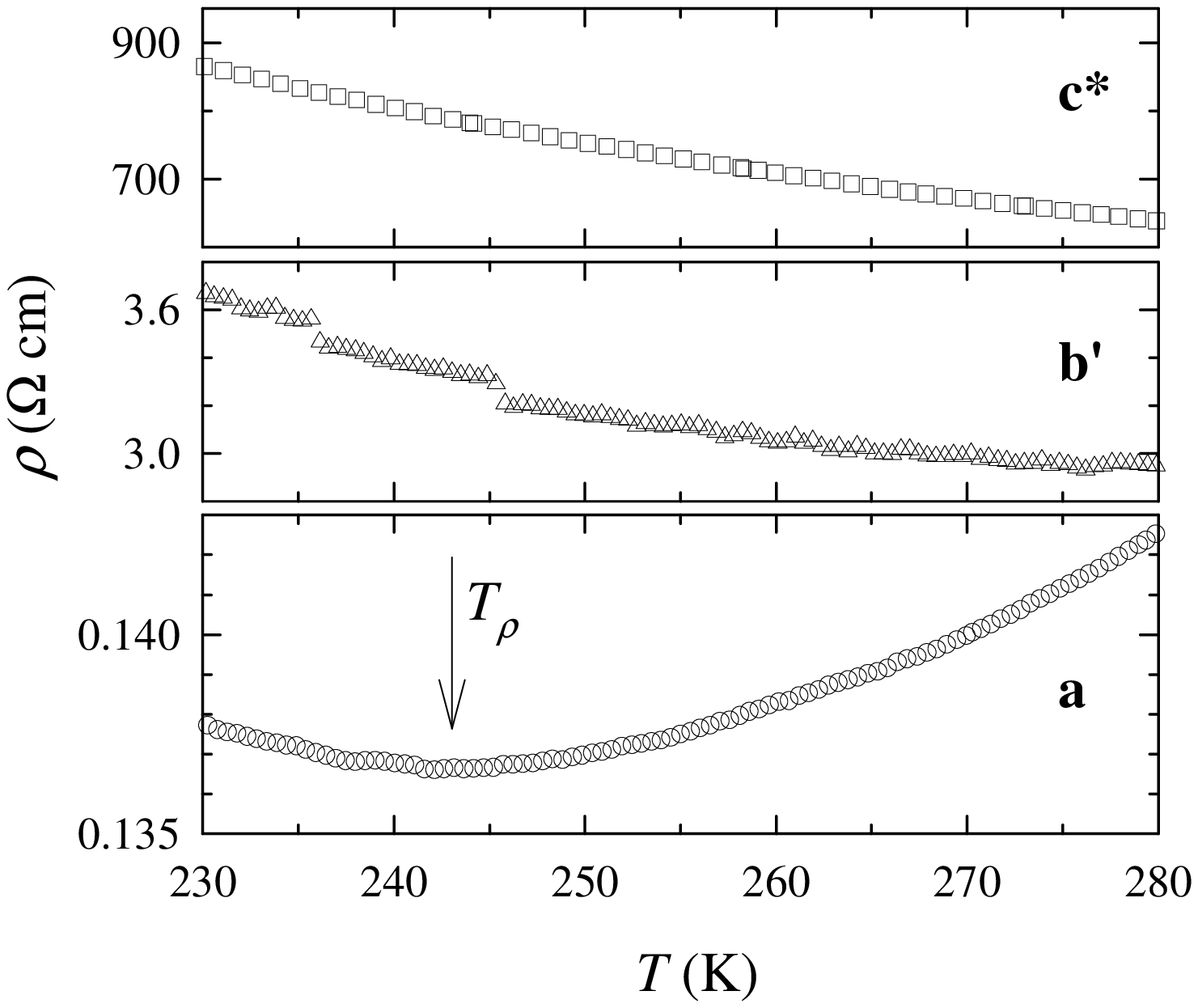}
\caption{\label{fig:3} Temperature dependencies of the resistivity along the
\va, \vb\ and \vc\ axes for \TAsF\ in the high-temperature region
($T>230\,$K).}
\end{figure}
room temperature conductivity values for $\sigma_{\mva}$ (\ja),
$\sigma_{\mv{b}}$ (\jb) and $\sigma_{\mv{c}}$ (\jc) are $(15\pm 5)$\Omcminv,
$(0.3\pm 0.1)$\Omcminv\ and $(1.5\pm 0.5)\times10^{-3}$\Omcminv, respectively.
The $\rhoa(T)$ results are in good agreement with the previously published
data,\cite{Laversanne84-5,Jacobsen82-16} while $\rhob(T)$ and $\rhoc(T)$ were
not measured up to now. The resistivity ($\rho$ vs.\ $T$) data above 230 K  are
shown in more details in \myreffig{fig:3}. One can observe the pronounced
difference in temperature dependencies of resistivities for three different
directions. The metallic-like behavior is found only for the highest
conductivity direction \va, above $240\,$K, and it can be described by a
$\rhoa\propto T^{0.75}$ law. Below a broad minimum around $T_\rho\approx
240\,$K, the resistivity $\rhoa$ increases on cooling and this is attributed to
the continuous opening of a charge gap: below $240\,$K, $\rhoa(T)$ can be
analyzed using a phenomenological law for a semiconductor $\rhoa(T) =
\rho_{\text{min}}\exp[\Delta_\mva/k_BT]$. In this  expression all the thermal
evolution of the resistivity $\rhoa(T)$ is included in the function
$\Delta_\mva(T)$, defined as a temperature-dependent energy gap, while the
prefactor $\rho_{\text{min}}$ is determined as $\rho(T_\rho)$ i.e.\ the
resistivity value just above $T_\rho\approx 240\,$K. $\Delta_\mva(T)$ is zero
at $T_\rho$ and its value gradually rises to almost $400\,$K down to \TCO. At
the charge-ordering transition \TCO\ there is the change in slope, and a steep
increase of the conductivity gap $\Delta$ is observed, yielding
$\Delta_\mva(\,T<\mTCO)\approx 480\,$K (see inset of \myreffig{fig:2}). \TCO\
is explained as the temperature where a spontaneous charge disproportionation,
called charge ordering (CO), occurs.

On the other hand, there are no indications of a metallic-like behavior along
neither of the two perpendicular directions (along \vb\ or \vc\ axis). Below
room temperature and in the temperature region where $\rhoa(T)$ is
metallic-like, for both \vb\ and \vc\ axis we have found that the resistivity
increase with decreasing temperature (\myreffig{fig:3}). These data could be
fitted to $\rhob(T)\propto T^{-0.5}$ and $\rhoc(T)\propto T^{-1.27}$ at high
temperatures.  Note also that there is no change in slope for neither
$\rhob(T)$ nor $\rhoc(T)$ around $T_\rho$, i.e.\ at the temperature where the
resistivity minimum is observed for the \va\ direction. Instead, for the \vb\
and \vc\ axes the resistivity increases exponentially down to $\mTCO\approx
100\,$K, and the respective conductivity gaps achieve similar values as for the
\va\ axis: $\Delta_\mvb=(350 \pm 50)\,$K and $\Delta_\mvc=(400 \pm 50)\,$K.
Below \TCO\ there is a sharper change of slope than for \va\ axis, indicating
an increased conductivity gap value $\Delta_\mvb(T<\mTCO) \approx
\Delta_\mvc(T<\mTCO) \approx 600\,$K.

\begin{figure}
\includegraphics*[scale=0.58]{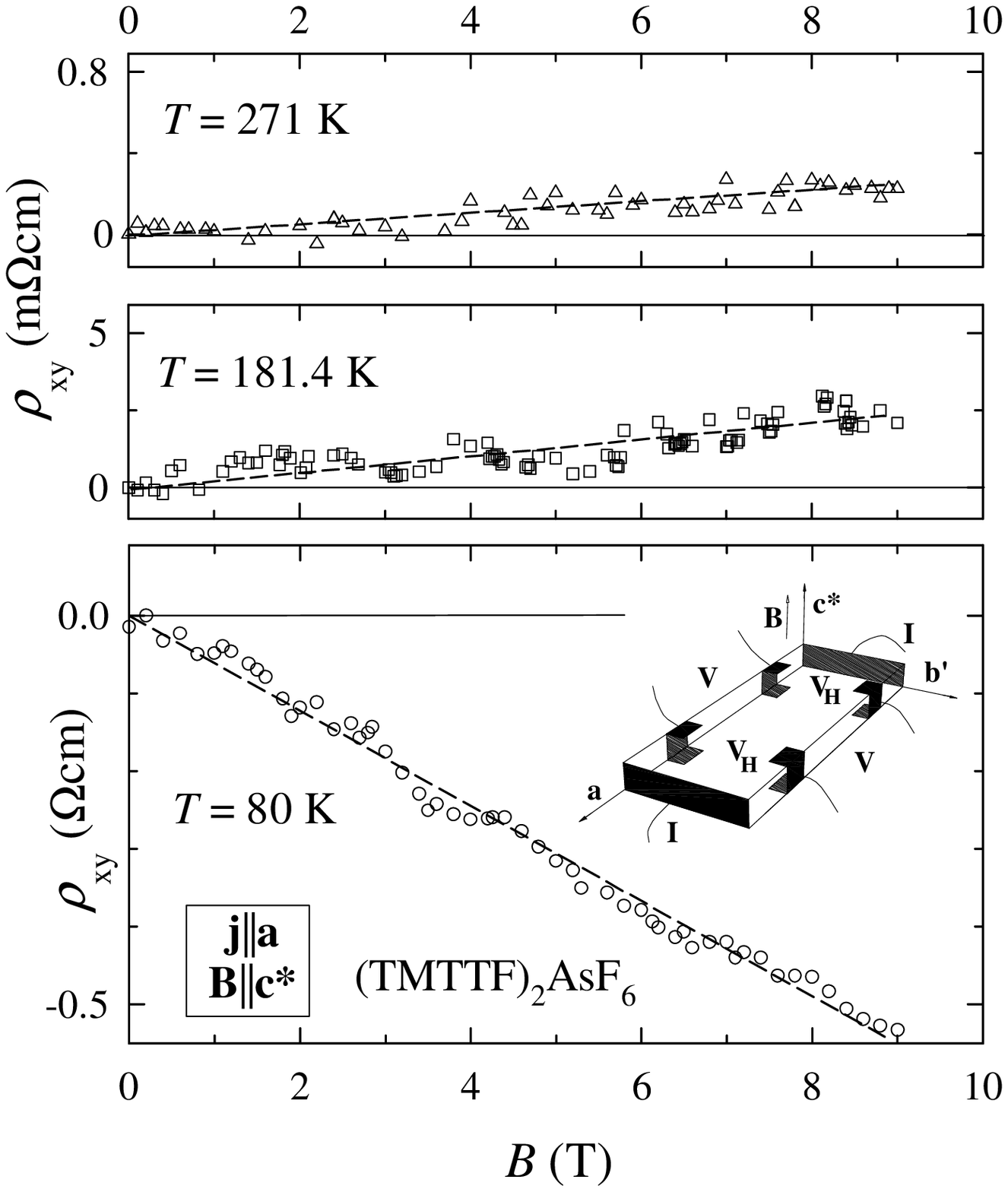}
\caption{\label{fig:1} Magnetic field dependence of the Hall resistivity of
\TAsF\ for several fixed temperatures. Also shown is the sample geometry (\ja,
\Bc) and the arrangement of contacts.}
\end{figure}

To our knowledge, no Hall experiments have been performed in any representative
of the \TX-family up to now. The magnetic field dependence of the Hall
resistivity  of \TAsF\ is shown in Fig.~\ref{fig:1} for three representative
temperatures. These data confirm that the Hall resistivity is linear with field
up to 9 T in the whole temperature interval we have investigated.

\begin{figure}
\includegraphics*[scale=0.58]{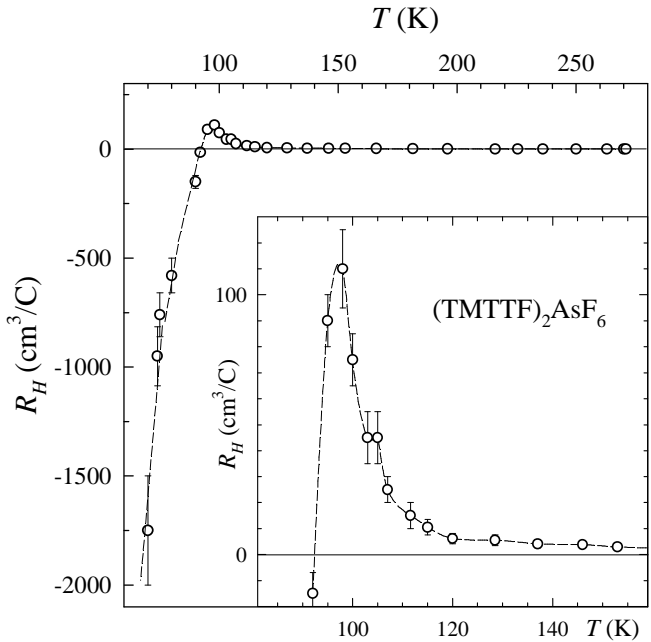}
\caption{\label{fig:4} Temperature dependence of the Hall coefficient $R_H$ of
\TAsF\ for $70\,\mathrm{K}<T<300\,\mathrm{K}$. For each temperature the Hall
coefficient values were determined as the average of the several measurements
taken during the cooling and heating cycles. Inset: $R_H(T)$ around \TCO.}
\end{figure}

Figure \ref{fig:4} shows the temperature dependence of the Hall coefficient
$R_H$ of \TAsF\ for $70\,\mathrm{K}<T<300\,\mathrm{K}$. (We note here that the
results for another crystal from the same batch showed similar behavior). The
inset of \myreffig{fig:4} shows (in more details) the same results around the
maximum in $R_H$. The Hall coefficient is positive from room temperature down
to about $95\,$K where it changes sign; then its (negative) value rapidly
increases with further temperature decrease. Note that the change in sign
occurs around \TCO, i.e.\ at the temperature where an increase of the
conductivity gap $\Delta$ is found in resistivity measurements.

\section{Discussion}
\subsection{Resistivity}
Let us start by pointing out that it is for the first time that in a member of
the \CX-salt family a metallic-like temperature variation of the resistivity at
high temperatures  has been found for \va\ axis only. Although only poorly
investigated up to now, for the Se-compounds the \vb-axis resistivity seems to
follows also a metallic-like decrease upon lowering the temperature, while for
\vc\ axis the results varied -- from metallic to semiconducting-like, depending
on $X$.\cite{Bojana03-14,Bechgaard80-17,Cooper86-18,Dressel05-19} On the other
hand, as far as we know, there are no published data on detailed resistivity
anisotropy measurements for the sulphur analogs \TX.

The simplest model of the FL electron transport in metals is the Drude theory,
where all the relaxation processes are described by a single relaxation time
$\tau$.\cite{Dressel02-20} In first approximation, lattice and electronic
interactions only renormalize the relevant quantities, leading to some
effective carrier mass, for instance. No matter how strong the anisotropy of
the parameters is, the underlying transport mechanisms should be similar in all
three directions leading to a similar dependence on temperature. Consequently,
this model cannot describe the obvious differences in the temperature
dependences of $\rhoa$, $\rhob$ and $\rhoc$, and therefore the resistivity
measurements provide evidence against a conventional FL picture. In particular,
the fact that the metallic-like behavior is found at high temperature for
$\rhoa$ only (for $T>240\,$K), further supports this opinion.

Within the model of a Luttinger liquid the transport properties were also
calculated for a system of weakly coupled chains, resulting in quantities like
the in-plane conductivity ($\sigma_\|$), inter-plane conductivity
($\sigma_\perp$) and Hall effect.\cite{Georges00-21,Lopatin01-22} It was found
that the inter-chain hopping is responsible for the metallic character of the
\CX\ compounds, which otherwise would be Mott insulators. Namely, the
interchain hopping can be viewed as an effective doping, leading to deviations
from the commensurate filling, which would be insulating. For longitudinal
($\rho_\|$) and transverse ($\rho_\perp$) resistivity, the temperature power
law was calculated, giving $\rho_\|(T)\propto T^{16K_\rho-3}$ and
$\rho_\perp(T)\propto T^{1-2\alpha}$. $K_\rho$ is the LL exponent controlling
the decay of all correlation functions ($K_\rho=1$ corresponds to
non-interacting electrons and $K_\rho<0.25$ is the condition upon which the
$1/4$ filled umklapp process becomes relevant) and $\alpha = (K_\rho +
1/K_\rho)/4-1/2$ is the Fermi surface exponent.\cite{Giamarchi04}

The large pressure coefficient of the conductivity in the metallic regime for
\CX\ salts ($\delta \ln\sigma/\delta P$ up to 25\% kbar$^{-1}$ around room
temperature) raises a problem for the comparison of experimental data with
theory, which usually computes constant-volume temperature
dependencies.\cite{Jerome82-23} Namely, it is known that in most organic
conductors much of the temperature dependence of the conductivity, at high
temperatures, arises from the thermal expansion. As a consequence, the constant
pressure data show usually different  temperature dependencies than
constant-volume data. Note that all the data presented in \myreffig{fig:2} are
measured at ambient pressure and usually considered as the constant-pressure
data. Therefore, the conversion from the constant-pressure $\rho(T)$ data in
\myreffig{fig:2} to the constant-volume $\rho^{(V)}(T)$ data has to be
performed, in order to compare the experimental results with theoretical
models. This conversion procedure was performed for \PF, where the unit cell,
at $4.2\,$K and at ambient pressure, was taken as the reference unit
cell.\cite{Jerome94-4} Here, the thermal expansion, the compressibility data
and transport data under various pressures would have to be taken into account.
Since we do not know these data for \TX\ compounds, we have followed the same
procedure as it was performed for \SX\ isostructural compounds. However, \TX\
salts have an approximately 5\% smaller unit-cell volume, about 2\% smaller
lattice parameters and about 30\% higher $t_\|/t_\perp$ ratio at $300\,$K
(calculated using the tight-binding scheme\cite{IshiguroBook98-1,Grant83-24})
than \SX\ salts, and therefore the conversion procedure has to be taken with
some precaution. We have used the expression $\rho^{(V)}_{\mva,\mvb} =
\rho_{a,b}/[1+(\delta \ln\sigma/\delta P)P]$, where $P$ is the pressure that
must be applied at given $T$ in order to restore the reference volume. We have
exploited the $P$ values from Ref.\ \onlinecite{Jerome94-4} and used in Ref.\
\onlinecite{Bojana03-14}, and we have taken $\delta \ln\sigma/\delta P=0.10$
(i.e.\ the same value as in Ref.\ \onlinecite{Dressel05-19}). For
$\rho^{(V)}_c$ the corrections were not made because the data for \vc\ axis
resistivity in \PF\ were calculated differently and $P$ values are not
available.\cite{Moser98-25}

\begin{figure}
\includegraphics*[scale=0.58]{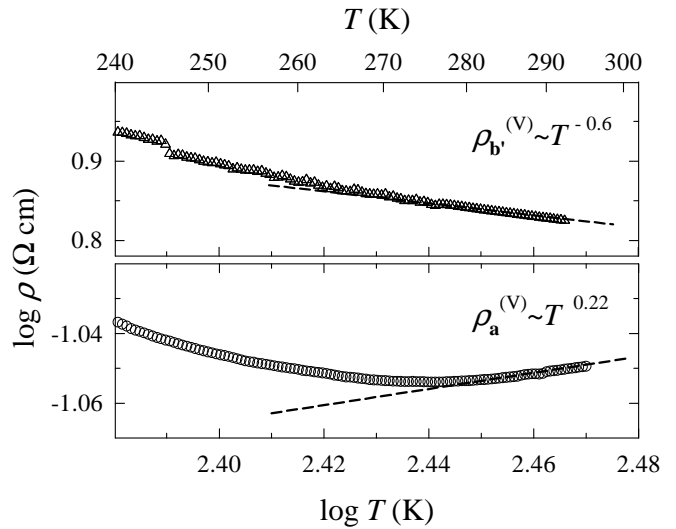}
\caption{\label{fig:5} Temperature dependencies of the constant-volume
resistivities $\rhoa^{(V)}$ and $\rhob^{(V)}$ (see text for the details of the
conversion procedure).}
\end{figure}

In the case of the \va-axis resistivity of \PF, the conversion from
experimentally obtained constant-pressure behavior $\rhoa(T)\propto T^{1.8}$
yielded an almost linear constant-volume dependence $\rhoa^{(V)}(T)\propto
T^{0.85}$ (Ref.\ \onlinecite{Jerome94-4}). Nearly linear $\rhoa$ vs.\ $T$ is
expected within existing FL and LL theoretical models (in this case for
$K_\rho\approx 0.23$ obtained from optical
measurements).\cite{Georges00-21,Zheleznyak99-26} Our experimental \va-axis
resistivity results for \TAsF\ show that already under constant-pressure
condition we obtain a smoother temperature dependence $\rhoa(T)\propto
T^{0.75}$ (for $T>260\,$K), i.e.\ even without conversion it is rather close to
a linear temperature dependence (the corresponding $K_\rho$ has about the same
values as for \PF). Nevertheless, the conversion procedure has to be performed
and this yields the constant-volume temperature dependencies
$\rhoa^{(V)}(T)\propto T^{0.22}$ and $\rhob^{(V)}(T)\propto T^{-0.6}$ (as shown
in \myreffig{fig:5}). From the obtained exponent values we found
$K_\rho=0.20$ and $\alpha=0.8$. Our $K_\rho$ is lower than 0.23 calculated for
\SX\ compounds: although this difference could be the consequence of a crude
approximation in the conversion procedure, it is worth noting that a somewhat
lower $K_\rho$ value is expected from the high-resolution photoemission
experiments \cite{Dardel93-27} corroborated with NMR measurements as
well.\cite{Wzietek93-28} Namely, it was suggested that $K_\rho$ increases when
going from (TMTTF)$_2$PF$_6$ to \PF, leading to a decrease of the $4k_F$
fluctuations and, correlatively, to a decrease of the charge localization
temperature. Up to now the measurements of dc transport along the
transverseaxis performed in \SX\ analogs were not fully understood
theoretically from the LL picture, since the model was compared with \va\ and
\vc\ axis resistivity results. The \vb\ axis data have shown anomalous
exponents and were considered as a crossover between the two
regimes.\cite{Dressel05-19,Moser98-25} Here, we emphasize that our \TAsF\
resistivity results show, for the first time, the agreement with LL model for
\va\ and \vek{b}\ axis. (We are neglecting \vek{c}\ axis because $t_b\gg t_c$.)

Summarizing this part of our results, we point out that the metallic-like
behavior is found at high temperature ($T>240\,$K) for $\rhoa$ only, which
evidences against a conventional FL picture. Furthermore, the power law of the
temperature dependence for the longitudinal and transverse resistivity proposed
in the LL model does agree with our $\rhoa^{(V)}(T)$ and $\rhob^{(V)}(T)$
experimental results for $T>260\,$K, with $K_\rho=0.20$, favoring the LL
description.

Our resistivity results for lower temperatures are consistent with those
already published,\cite{Laversanne84-5} showing a broad minimum around
$T_\rho\approx 240\,$K for $\rhoa(T)$ and an increase below $T_\rho$. No change
in the magnetic susceptibility (indicating the spin-charge separation) or the
change in the structure has been observed at $T_\rho$. It is suggested that the
presence of the anomaly of such description indicates that the electronic
system is highly correlated. This behavior is a consequence of the continuous
opening of a charge gap closely connected to the increased Coulomb interaction
and dimerization. However, the results for \vb\ and \vc\ axis do not show any
change in slope around $T_\rho$ and have not been analyzed up to now. At
variance with ordinary semiconductors, only charge degrees of freedom are
thermally activated below $T_\rho$ whereas spin excitations remain gapless, a
feature typical of a Mott insulator.\cite{Bourbonnais99-6}

For $T<T_\rho$ the resistivity increases exponentially down to
$\mTCO\approx100\,$K where there is a change of slope (cf. Fig. \ref{fig:2}),
i.e.\ a steep increase of the conductivity gap $\Delta$: sharper changes in
\vb\ and \vc\ directions indicate an anisotropy of the $\Delta$ values. \TCO\
is the temperature where a spontaneous charge disproportionation, or charge
ordering, occurs, dividing the molecules into two nonequivalent species as
verified using 1D and 2D NMR spectroscopy. \cite{Chow00-8} At ambient
temperature the spectra are characteristic of nuclei in equivalent molecules.
Below a continuous charge-ordering transition temperature \TCO, there is
evidence for two non-equivalent molecules with unequal electron densities. This
is supported by infrared spectroscopy of the EMV-coupled $A_g(\nu_3)$
intramolecular vibration.\cite{Dumm04} Below \TCO\ the mode splits in two
indicating a charge dispropotionation of $\pm 0.13e$. The absence of an
associated magnetic anomaly \cite{Dumm05} indicates that only the charge
degrees of freedom are involved and the lack of evidence for a structural
anomaly suggests that charge-lattice coupling is too weak to drive the
transition. EXAFS experiments were performed below and above \TCO, in order to
check a possible displacement of the anion coupled to the charge
order.\cite{Ravy04-29} No significant difference was observed between the
spectra, indicating that the displacement of the anion, if any, is less than
$0.05\,$\AA. The nature of possible phase transition was clarified by
discoveries of the huge dielectric anomaly, and the phase transition was
interpreted as the one to the ferroelectric (FE) state. It was also suggested
that the FE transition is triggered by the uniform shift of ions yielding a
macroscopic FE polarization which is gigantically amplified by the charge
disproportionation at the molecular stack.\cite{Brazovskii04-30} The fact that
the increase in $\Delta$ is anisotropic or somehow higher along the \vb\ and
\vc\ axes than along the \va\ axis may further support this suggestion.

\subsection{Hall Effect}
\begin{figure}
\includegraphics*[scale=0.58]{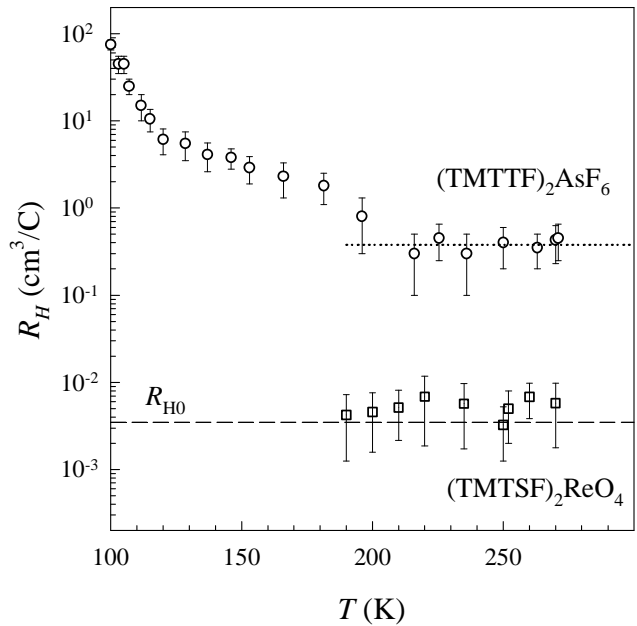}
\caption{\label{fig:7} Temperature dependencies of the Hall coefficient $R_H$
for \TAsF\ (circles) and (TMTSF)$_2$ReO$_4$ (squares). Above $200\,$K $R_H$ of
\TAsF\ has an almost constant value $0.38\,$cm$^3$C$^{-1}$ (dotted line). The
dashed line represents the value $R_{H0} = 3.31\times 10^{-3}\,$cm$^3$C$^{-1}$
obtained for \TAsF\ and the carrier concentration of one hole/unit cell ($n =
1.4\times 10^{21}\,$cm$^{-3}$).}
\end{figure}

The Hall constant $R_H$ of \TAsF\ for $T>\mTCO \approx 100\,$K is positive,
hole-like and temperature-dependent up to $\approx 200\,$K. The decrease in
$R_H(T)$ with increasing temperature (for $\mTCO < T < 200\,$K) is consistent
with the continuous charge localization already pointed out when describing the
$\rhoa(T)$-behavior below $T_\rho$. Above $200\,$K, $R_H$ is almost constant
with the value around $0.38\,$cm$^3$C$^{-1}$ (dotted line, \myreffig{fig:7}).

The confirmation of one-dimensionality and the applicability of LL model at
high temperatures ($T>200\,$K) to the transport properties in \TAsF\ should be
valid for the Hall effect as well. As already mentioned in the introduction, it
was concluded from the optical conductivity results for \CX\ salts
\cite{Vescoli98-10} (considered as the strong evidence for LL behavior) that
all the dc transport is due to a very low concentration of carriers and that
needed to be verified experimentally.

Theoretical analysis shows that the Hall coefficient $R_H$ of weakly coupled 1D
Luttinger chains does not depend on frequency and temperature, and it is given
by a simple formula corresponding to the non-interacting fermions, i.e.\ $R_H =
R_{H0}\approx 1/ne$ (where $n$ is the concentration of carriers). More
precisely for quasi-1D systems, using the tight-binding dispersion along the
chains, the obtained Hall constant is $R_H=(1/ne)(k_Fa/\tan k_Fa)$ (where $e$
is the electric charge and $k_Fa =\pi/4$).\cite{Cooper77-31,Maki90-32} The
carrier density for \CX\ salts of $1\,$hole/unit cell gives for \TAsF\
$n=1.48\times 10^{21}\,$cm$^{-3}$, and $R_{H0} = 3.31\times
10^{-3}\,$cm$^3$C$^{-1}$: this is represented as a dashed line in
\myreffig{fig:7}, and this value is about 100 times lower than the
experimentally obtained one.
In other words, the carrier concentration deduced from the experiment is two
orders of magnitude smaller than expected by simple counting arguments,
indicating that only a fraction of carriers (about 1\%) contribute to the
transport. Our findings are in good agreement with the optical investigations
as well as with the LL theoretical predictions based on doped Hubbard chains.
They strongly evidence the existence of a high temperature regime (above
$200\,$K) in \TX\ salts, where 1D physics dominates, i.e.\ where 1D LL features
appear in the transport properties. Here, we would like to repeat that all Hall
effect data published up to now were performed for \SX\ systems only, and in
all those experiments the experimentally obtained $R_H$ values indicated that
all the carriers contribute to the metallic-like transport. Consequently, such
data disagreed with the conclusions from the optical conductivity measurements.
Moreover, comparing our Hall effect results for (TMTSF)$_2$ReO$_4$ (Ref.\
\onlinecite{Bojana03-14}) and \TAsF\ (\myreffig{fig:7}) we find indications
that the high temperature transport properties in \CX\ salts change their
behavior when going from a LL (for $C=\mathrm{S}$) to a FL (for
$C=\mathrm{Se}$) regime.

The Hall coefficient exhibits an abrupt change in its temperature
dependence around \TCO\ (cf.\ inset of \myreffig{fig:4}): $R_H$
changes its sign and its value increases rapidly with the further
temperature decrease. While the increase of $R_H(T)$ with the
temperature decrease below $T_\rho$ is expected because of a
gradual localization of charge carriers, the abrupt change, i.e.\
the change of sign, around $\mTCO\approx 100\,$K is surprising
since it occurs in the semiconducting state. Note, the change in
sign of the Hall coefficient
was found at the SDW transition in \PF\ (Refs.\ %
\onlinecite{Mihaly00-12,Moser00-13}), as well as at the anion
ordering transition in (TMTSF)$_2$ReO$_4$.\cite{Bojana03-14} In
both cases the phase transition is also a metal-insulator
transition, connected with the opening of the energy gap and
consequently the change in transport mechanism from metallic-like
to semiconducting one. On the other hand, the stronger temperature
dependence of $R_H$ close to the phase transition followed by the
change of sign (indicating that the holes are participating in the
electrical transport above and electrons below phase transition)
is a common feature in semiconductors.\cite{Beer_hall-33} As the
Hall fields created by electrons and holes are opposing each
other, it is apparent that the galvanomagnetic effects can have
unusually strong temperature variations in regions where the
resultant Hall field is nearly null, when the relative
electron-hole population is temperature dependent. Below the CO
transition $R_H(T)$ shows the rapid increase with decreasing
temperature, i.e.\ the Hall coefficient is activated as expected
for semiconductor. Moreover, the activation energy agrees well
with that obtained from the resistivity data. The question arises
why the dramatic change in $R_H$ does not occur in the temperature
range where the resistivity shows its minimum and the carriers
become localized, as  identified by the increase of the
conductivity gap. Instead, not much change is observed well below
$T_\rho$ until charge ordering sets in at the transition \TCO. We
propose the following interpretation: if one imagines that below
$T_\rho$ the gap opens gradually upon cooling, the increase in
$R_H(T)$ indicates a decrease in number of carriers (holes) as
long as $\Delta(T)\le k_B T$. In the semiconducting phase around
\TCO\ where $\Delta(T)>k_B T$, a transition occurs in which the
carriers condense into an insulating charge-ordered state, i.e.\
our results reflect the exponential freezing out of the carriers
below \TCO. In line with these, the abrupt change in $R_H(T)$
around \TCO, where an increase of the conductivity gap $\Delta(T)$
occurs, demonstrates a phase transition to a charge ordered
state.\vspace{0.9cm}

\section{Conclusion}

We have found clear evidence that the transport properties of \TAsF\ at high
temperatures ($T>200\,$K) can be understood in terms of the Luttinger-liquid
theory which describes 1D systems. This conclusion is based on the
metallic-like resistivity along the \va-axis only for $T>T_\rho$, as well as on
a good agreement with the power laws predicted for $\rhoa$ ($\rho_\|$) and
$\rhob$ ($\rho_\perp$) from existing LL model. Further support comes from the
constant value of the Hall coefficient in the same temperature region, showing
a strong reduction in number of carriers participating in dc transport. By
comparing the results on \TAsF\ with those for \SX\ systems (where the LL power
laws for the resistivity yield a more complex picture, while $R_H$ has much
lower values) we can infer that the high temperature transport properties in
\CX\ salts change their behavior from a Luttinger liquid to a Fermi liquid when
going from $C=\mathrm{S}$ to Se.

At lower temperatures the resistivity data show a semiconducting
behavior for all three directions down to $\mTCO\approx100\,$K,
where a charge-ordering transition occurs. This is characterized
by a steep increase of the conductivity gap in the resistivity
measurements for all three directions, although somehow stronger
for the two axes perpendicular to the chains. This may be the
consequence of the uniform shift of ions yielding a macroscopic FE
polarization as predicted in analysis of dielectric measurements.
However, the fact that no significant change in structural
measurements was found below \TCO\ weakens this suggestion. On the
other hand, the steep increase of the conductivity gap below \TCO\
in resistivity manifests itself as an abrupt change in the
$R_H(T)$ behavior: the change of sign and the rapid increasing of
its value with further temperature decrease is a strong indication
that carriers condense into an insulating charge-ordered state.

\begin{acknowledgments}
We thank G.\ Untereiner for the crystal growth and sample
preparation. The work was supported by the Croatian Ministry of
Science, Education and Sports, grant 0119251 and 0035015 and the
Deutche Forschungsgemeinschaft (DFG).
\end{acknowledgments}

\end{document}